\documentclass[12pt]{cernart}
\usepackage{graphicx}
\usepackage{here}
\begin{document}
\begin{flushright}
CERN-DG/98-3534 \\
hep-ph/9812301 \\
\end{flushright}
\title{INELASTIC SUM RULES}
\author{C.H. Llewellyn Smith}

\begin{center}
{\it Talk given at the Sid Drell Symposium \\
SLAC, Stanford, California, July 31st, 1998}\\
{~} \\
{\bf Abstract}\\
{~} \\
The history and present status of several sum rules for deep-inelastic
lepton scattering are reviewed, with particular attention to the discovery
of scaling, partons, quarks and QCD. Two outstanding issues are then
discussed in more detail: the singlet (Ellis-Jaffe) nucleon spin
sum rule and the Drell-Hearn-Gerasimov-Iddings sum rule. \\
{~} \\
\end{center}

\section{Introduction}

It is a great pleasure to be here today to honour Sid Drell,
and a great honour for me to have been invited to speak.  I
first heard of Sid in October 1964 when, as a beginning 21-year-old
graduate student, I
went to see my supervisor Dick Dalitz for the first time.  Dick told me
that he would like
me to learn to calculate as quickly as possible, adding {\em `There is a
new book by Bjorken and Drell: I have not seen it yet myself,
but those characters would write a good book'}.  He was
absolutely right, and it was my bible for many years to
come.

I first met Sid in September 1970 when I arrived here from CERN
as a postdoc.  I had the office next to his (which is the
office occupied today by David Leith).  Sid operated an
open-door policy, and everyone was allowed to gather in his
office, attracted by his loud voice, and participate in any
discussion that happened to be going on.  All of us postdocs
and students learned a great deal of physics from those
discussions, and we were also able to participate vicariously
in the non-secret parts of Sid's Washington life.

Towards the end of my two years at SLAC, Sid went away for a
short sabbatical and he and Harriet kindly allowed us to move
into their house.  This experience gave me a lot of insight
into the Drell family and into their three characteristically
named cats --- Harvard, Princeton and Yale.

 Subsequently, our paths crossed many times, for example when
Sid came on sabbatical to Oxford.  Most recently our world
lines became intertangled through the HEPAP Panel (the so-called
Drell Panel) that Sid chaired in 1994, which helped start what
I hope is the dawn of a new age of global collaboration in
particle physics.  On behalf of all those who are involved with
the LHC, in over 40 countries including the 19 Member States
and the USA, I would like to take this opportunity to thank Sid
for the statesman-like report that his Panel produced.

The first part of this talk will comprise a historical
overview, together with a comparison with recent data, of
various sum rules, with emphasis on the crucial role that they
played in the discovery of the scaling phenomena, partons,
quarks and eventually QCD.  The bona fide sum rules are built
into the parton distributions that are used nowadays to fit the
data, so we can anticipate straightaway from the consistency
of the fits that these sum rules all work.

In the second part of my talk, I shall deal with two
outstanding issues: the singlet (Ellis--Jaffe) nucleon spin sum
rule and the Drell--Hearn--Gerasimov--Iddings sum rule, and the
connection between these sum rules.  Except in the final
section, there will be  little explicit mention of Sid's work in
this talk (it is discussed more extensively in the preceding
talk by Bob Jaffe and the following talk by Tung-Mo Yan), but
his virtual presence pervades what I have to say, and --- I believe ---
is strongly implicit in everything that has come from SLAC,
  experimental as well as theoretical.

\section{Historical Overview}

I would like to start by picking up on some remarks of Bob
Jaffe about the climate in the mid-1960s.  At that time,
electron and photon interactions were a minority interest, and
it was widely believed that the way ahead was to study proton
collisions.  Indeed, Sid's realization that SLAC
would be a good source of pion beams was an important factor in
legitimizing the construction of SLAC in the eyes of many
people.  There was almost no interest in inelastic lepton
scattering, and only one paper was submitted on this subject at
the 1967 SLAC Electron--Photon Conference.  As Bob
Jaffe has recalled, the focus was on nuclear democracy and bootstraps,
and quarks were generally considered to be (at best) a
heuristic tool.  Chew's statement, in 1961, that  {\em `I believe the
conventional
association of fields with strongly interacting particles to be empty. I do
not have firm
convictions about leptons or photons ... field theory ..., like an old
soldier, is destined
not to die but just to fade away.'} was,  I believe, widely accepted.

My story begins in 1965 with the Adler sum rule \cite{lleweref1} (although it
could have started in 1964 with the Adler--Weissberger relation,
between $g_A$ and $\sigma_{\pi N}$, which can be derived from the Adler sum
rule by taking the limit $q^2 \rightarrow 0$  and using Adler's PCAC forward
neutrino theorem).  In modern notation, which was not used at
the time, the Adler sum rule reads:

\begin{equation}
\int^1_0 \left({F_2^{\nu n} (x, q^2) - F_2^{\nu p} (x, q^2)}\right)
\frac{dx}{x} =2\;.
\end{equation}
This sum rule can be, and was, also written in the form:

\begin{equation}
\lim_{E\rightarrow \infty} \frac{d\sigma^{\nu n}}{dq^2} -
\frac{d\sigma^{\nu p}}{dq^2} =
\frac{G^2}{\pi}\;.
\end{equation}

I will return to the parton interpretation of these results (in
terms of the difference of the number of isospin-up and
isospin-down constituents in the nucleon) later, but I would
like to emphasize that it is {\em not} a parton sum rule, and is
indeed exact  for all $q^2$ (it can be derived by integrating the component
$W_{00}$ of the deep inelastic scattering tensor $W_{\mu \nu}$ over $q_0$,
taking
the limit $|\vec p|\rightarrow \infty$ and assuming that this limit can be
exchanged
with the integral, or alternatively by other manipulations
which depend on assuming an unsubtracted dispersion relation
for the structure function $W_2$).  The forms (1) and (2) are
highly suggestive of scaling and of scattering off point-like
constituents in the nucleon, as was --- uniquely --- recognized by
Bjorken.

Bj's first contribution to the subject was in his classic 1966
paper \cite{lleweref2}, to which Bob Jaffe has referred, in which --- among
many
other things --- he derived the polarization sum rule:

\begin{equation}
\int \left({{g^p_1} (x, q^2) - g^n_1 (x, q^2)}\right) dx = \frac{1}{6}
\left|{\frac{g_A}{g_V}}\right|
\left({1-\frac{\alpha_s}{\pi} }\right)\;,
\end{equation}
where I have included the leading-order QCD correction (which
obviously was not known at the time, but is now known up to and
including order of $\alpha_s^3$).  I will discuss the parton
interpretation and derivation of this equation later.  It was
derived at the time by Bj by taking the `Bj limit' $q_0 \rightarrow i \infty$
followed by the limit $|\vec p|\rightarrow \infty$.  This procedure was
later used by
Bj to `derive' scaling.  Assuming scaling, it leads to the
generic Cornwall--Norton  sum rules \cite{lleweref3}, of which one example is:
	\begin{equation}
2 \int F^\pm _1 \frac{d\omega}{\omega^{n+2} } = \lim_{P_0\rightarrow \infty} i
\left({\frac{i}{2P_0}}\right)^{n+1} \int \langle p| \left[{\frac{\partial^n
J^+}{\partial t^n} (\vec
x,0),\,J^- (0)}\right] |p\rangle d^3 x
\end{equation}
and similar sum rules for the other structure functions.  These
generic sum rules were used to derive most of the specific sum
rules discussed below.

In his 1966 paper, Bj dismissed his sum rule for $g_1$ as
 {\em  `worthless'}, but it has now been tested to great accuracy. The
latest SMC analysis \cite{lleweref4} of the Bj sum rule, carried out at
$q^2 =5$~GeV$^2$
and including next-to-leading-order QCD corrections, implies:
$$\frac{g_A}{g_V} = 1.15\pm 0.03^{+0.07\,+0.14}_{-0.06\,-0.04}\;,$$
where the last error is theoretical, compared to the Particle Data Group
value of
$1.2601\pm 0.0025$.

From his {\em `worthless'} sum rule Bj salvaged an inequality for
unpolarized scattering which led to the conclusion that
{\em `inelastic scattering is large ... comparable to scattering off
point-like charges'}.
 In his 1966 paper, Bj used the same techniques to derive the
fact that the electron--positron annihilation cross section
should scale, and stated that

{\em `The idea that the total hadronic
yield from colliding $[e^+e^-]$ beams should be approximately the
same as the $\mu^+\mu^-$ yield is folklore$^{33}$. 33) B. Richter (private
communication).'}
 Pretty select folk (perhaps only Bj and Richter!), and I
believe that the prevalent view was that the cross section
should vanish rapidly, damped by form factors.  In the same
paper, Bj also argued that the total neutrino cross section should
rise linearly with energy.

1966 also witnessed the publication of the Drell--Hearn sum
rule \cite{lleweref5}.  It had in fact been discovered independently by
Gerasimov and published \cite{lleweref6} in Russian in 1965, but the English
translation only appeared in 1966.  It also appears in a 1965
Physical  Review paper by Iddings on the nucleon polarizability
contribution to the hyperfine structure of hydrogen \cite{lleweref7}, but
Iddings did not suggest testing the sum rule or discuss its
implications.  I will return to this sum rule at the end of this
talk.

The most important contribution in 1967 was, in my opinion,
 made by Bj  at the SLAC Electron--Photon Conference \cite{lleweref8} (see
also his Varenna Summer School lectures \cite{lleweref9}).  In his talk, Bj
developed a description of the Adler sum rule in terms of
incoherent scattering off point-like constituents.  After
quoting Eq.~(2) above, Bj stated that {\em `This result would
also be true were the nucleon a point-like object, because the
derivation is a general derivation.  Therefore the difference
of these two cross sections is a point-like cross section, and
it is big'}.  He goes on to suggest an interpretation, as
follows: {\em `We assume that the nucleon is built of some kind of
point-like constituents which could be seen if you could really
look at it instantaneously in time ...  If we go to very large
energy and large $q^2$ ... we can expect that the scattering will
be incoherent from these point-like constituents.  Suppose ...
these point-like constituents had isospin one-half ... what the
sum rule says is simply $[N\uparrow] - [N\downarrow ] = 1$ for any
configuration
of constituents in the proton.  This gives a very simple-minded
picture of this process which may look a little better if you
really look at it, say, in the centre-of-mass of the lepton and
the incoming photon.  In this frame the proton is ...
contracted into a very thin pancake and the lepton scatters
essentially instantaneously in time from it in the high energy
limit.  Furthermore the proper motion of any of the
constituents inside the hadron is slowed down by time
deletion.  Provided one doesn't observe too carefully the final
energy of the lepton to avoid trouble with the uncertainty
principle, this process looks qualitatively like a good
measurement of the instantaneous distribution of matter or
charge inside the nucleon'}.

Bj then went on to apply similar ideas to the electromagnetic
cross section, reaching the conclusion that it also should be
big and point-like.  It is clear from these quotations that Bj
already had the essential ideas of the parton model in 1967,
although he did not make any explicit statement about scaling.

Bj's `derivation' of scaling was worked out in 1968, although
not published \cite{lleweref10} until 1969.  In his paper Bj says that {\em
`A more
physical approach into what is going on is, without question,
needed'}.  Thus Bj had partons without scaling in 1967, and
apparently scaling without partons in 1968!  Surely he must
have put these ideas together, even if he did not tell the
world.  I have put the word `derivation' in inverted commas
above, because it depended on formal manipulations, which,
although they appear to be valid in any renormalizable field
theory, do not in fact  completely survive the effects of
renormalization, as will be discussed below.

The first inelastic scattering results from SLAC were presented
by Panofsky \cite{lleweref11} at the Vienna conference in 1968.  Panofsky
reported that these cross sections {\em `are very large and decrease
much more slowly with momentum transfer than the elastic
scattering cross sections and the cross sections of the
specific resonance states ... therefore theoretical speculations
are focused on the possibility that these data might give
evidence on the behaviour of point-like, charged structures
within the nucleon ...  The apparent success of the
parametrization of the cross sections in the variable $\nu/q^2$ in
addition to the large cross section itself is at least
indicative that point-like interactions are becoming
involved'}.  Panofsky also discussed the experimental status of
a sum rule derived by  `Godfrey' who treated the proton in a non-relativistic
point quark model.  This was in fact a reference
to a paper \cite{lleweref12} by Gottfried, published in 1967 before Bj's
talk at
the SLAC conference, to which I now return.

Gottfried noted that in the {\em `breathtakingly crude'} na\"{\i}ve
three-quark model the second term in the following equation
vanishes for the proton (it also vanishes for the neutron, but
neutrons are not mentioned):
\begin{equation}
\sum_{i,j} Q_i Q_j \equiv \sum_i Q^2_i + \sum_{i\neq j} Q_i Q_j\;.
\end{equation}
Thus for any charge-weighted, flavour-independent, one-body
operator all correlations vanish, and therefore using the
closure approximation the following sum rule can be derived:
	\begin{equation}
\int_{\nu_0} W^{ep}_2 (\nu, q^2) d\nu = 1 - \frac{G_E^2 - q^2 G_M^2 /
4m^2}{1- q^2 /
4m^2}\;,
\end{equation}
where $\nu_0$ is the inelastic threshold
(the methods used to derive this sum rule are those that have
 long been used to derive sum rules in atomic and nuclear
physics, for example the sum rule \cite{lleweref13} derived in 1955 by
Drell and
Schwarz).  After observing that this sum rule appears to be
oversaturated in photoproduction (we now know that the
integral is actually infinite in the deep inelastic region),
Gottfried asked whether it was {\em `idiotic'}, and stated that if,
on the contrary there is some truth in it, one would want a
{\em `derivation that a well-educated person could believe'}.

In his talk at the 1967 SLAC conference Bj quoted Gottfried's
paper and stated that diffractive contributions should
presumably be excluded from the integral, which could be done
by taking the difference between protons and neutrons, leading
to the following result, in modern notation:
\begin{equation}
\int \left({F_2^{ep} (x, q^2)  - F_2^{en} (x, q^2)}\right) \frac{dx}{x} =
\frac{1}{3}\;.
\end{equation}

This result, which is generally known as the Gottfried sum
rule, is not respected by the data which give the value \cite{lleweref14}
$0.235\pm
0.026$.  In parton notation, the left-hand side can be written
\begin{equation}
\frac{1}{3} (n_u + n_{\bar u} -n_d - n_{\bar d}) = \frac{1}{3} +
\frac{2}{3} (n_{\bar
u}-n_{\bar d})\;,
\end{equation}
where the second expression follows  using isospin
conservation.  The fact that the data give a number less than
one-third implies that $n_{\bar d} > n_{\bar u}$  which is not implausible
if we
note that
 i) the proton can virtually dissociate into $\pi^+n$ in
which they are more anti-d's than anti-u's (as well of course
into $\pi^0p$ in which the numbers are equal), and ii) one might
expect the production of up anti-up pairs in the nucleon to be
suppressed relative to the production of down anti-down pairs
as a result of the Pauli principle.  Although not exact, the
Gottfried--Bj sum rule is very interesting as it was the first
result to give information about quark/parton charges.

In 1969 the parton model emerged from Bj and Feynman's
notebooks, and Bjorken and Paschos published \cite{lleweref15} an explicit
parton
model, with model functions for the quark distributions, but no
gluons.  1969 also witnessed the publication of the first of
the series of papers \cite{lleweref16} by Drell, Levy and Yan.  They studied a
canonical field theory of pions and nucleons, with a cut-off on
the transverse momentum applied in the frame in which target
particles are moving with infinite momentum.  This model
provided a very important laboratory for identifying those
processes, other than deep inelastic scattering, to which
parton ideas might apply.  It led not only to the classic
predictions for what is now known as the Drell--Yan process, but
also to the identification of most of the other processes which
we now know can  be consistently described by perturbative QCD.

In 1969 Callan and Gross published the first paper \cite{lleweref17} that
used sum rules
involving commutators of  a time derivative of a current with another
current [see Eq.~(3)
above] to make testable predictions depending on the  model that had been
used to derive
the commutator.  In particular, by studying the sum rules for moments of
the structure
function $F_2$ and
$F_1$ they found that:
\begin{eqnarray}
\frac{\sigma_L}{\sigma_T} &=& 0 - \mbox{quark model}\;,\cr
&=&\infty - \mbox{algebra of fields}\;.
\end{eqnarray}
This sum rule can easily be interpreted, using the parton model in
the Breit frame for absorbing a virtual photon by a
constituent, in terms of the ability of constituents of
different spin to absorb photons with different helicities, but the parton
interpretation
followed the formal derivation of the sum rule.  It is, of course, now
known that
$\sigma_L/\sigma_T$ is small, except at very small $x$, and that the value
given by QCD
is not exactly 0 but of order $\alpha_s$ due to scattering off
quark--anti-quark pairs with
non-zero transverse momentum produced from gluons.

In 1969 I got involved in the sum rule/parton business, which
by then had become relatively simple since essentially all the
pre-QCD techniques were well developed (use of the Bjorken
limit, the infinite momentum frame, current commutators and
commutators of currents with their time derivatives).  David
Gross arrived at CERN, where I was then a postdoc, on
sabbatical in early 1969 and gave a talk on the Callan--Gross
relation and other recent developments.  I was then working
with John Bell on shadowing of neutrino interactions in nuclei,
and as an exercise to see whether I had understood the
description in David's talk, I applied the Bjorken--Paschos
model to neutrino structure functions.  I showed the results to
David, who became very interested when he realized, looking at
my results, that there is an additional structure function ($F_3$)
in neutrino interactions.  He immediately suggested that we see
whether we could use the formal methods then in vogue to derive any result
for the moments of $F_3$. After
taking the appropriate limits, we derived on the blackboard the sum rule
\cite{lleweref18}:
\begin{equation}
\int F_3^{\nu N} \frac{dx}{2x} = 3\;.
\end{equation}

We  did not put the factor one-half on the left-hand
side and therefore had six on the right.  David's immediate
reaction was that sum rules give numbers like 1 or 2, but not
6, and that we must therefore have made a mistake.  After
checking the algebra and not finding any obvious error, I
decided to check the sum rule in the Bjorken--Paschos model.  It
was only after elaborately integrating over their explicit
models for quark distribution (which were expressed in a
complex way as sums over contributions from configurations with
different numbers of quark--anti-quark pairs) that I realized
that in fact the sum rule has to be true in any quark model
since it simply states that:
\begin{equation}
n_q - n_{\bar q} = 3
\end{equation}
because of baryon conservation.  This sum
rule obviously provides a critical test of the quark model.
There are important QCD corrections on the right-hand side,
which have now been calculated to order $\alpha_s^3$.  Including these
corrections, the final analysis of the CCFR  data gives \cite{lleweref19}:
$$\alpha_s (M_Z) = 0.114^{+0.005\,+0.007\,+0.004}_{-0.006\,-0.009\,-0.003}$$
where the last error is the result of an estimate of possible higher twist
contributions,
which  may be conservatively large.  This result is to be compared with the
values of
$0.120 \pm 0.003$, obtained from fits to precision electroweak data, and
$0.119\pm 0.004$
from fitting all the data.

It was pointed out during 1969 \cite{lleweref20} that,  although
non-trivial interaction
Hamiltonians were assumed, the formal manipulations used to derive scaling
and the
scaling-limit sum rules
    are all invalid in interacting field theory, and in particular are
spoiled by
logarithms of $q^2$.  Of course we now know that in QCD these
logarithms can be summed up to give corrections that vary as $\alpha_s \sim
\ln (q^2)^{-1}$, or powers of
ln$(q^2)$ given by anomalous dimensions.  At the time, many of us were
prepared to dismiss
the logarithms that appeared in perturbation theory on the basis
that the data seemed to support the underlying picture and that
nature must somehow behave in a smoother way than that
predicted by all the field theories then known.  A typical
reaction is that in my paper with David Gross, in which we
arrogantly state that {\em `In second-order perturbation theory all the
limits ... are
infinite .... [and] the sum  rules ... diverge.  In the case of
electron--nucleon
scattering this is contradicted by experiment. Thus the real world is less
divergent than
pertubation theory indicates'!}

 In late 1969, I decided to look for all sum rules that
could be derived in arbitrary quark-parton models (a relatively
simple exercise), and then see if they could all be `properly'
derived using the formal techniques discussed above.  The
result was the following two relations \cite{lleweref21}:
\begin{equation}
F_3^{\nu p} - F_3^{\nu n} = 12 (F_1^{ep} - F_1^{en})
\end{equation}
\begin{equation}
F_2^{ep} + F^{en}_2 \geq \frac{5}{18} (F_2^{\nu p} + F_2^{\nu n})
\end{equation}
where the second expression is an equality if there are no
strange quarks. Both relations were `derived' for all moments
of the structure functions from equations like (3) in quark models with
model interaction Hamiltonians.

Equation (13) is known to be rather well satisfied as an
equality.  I promoted it as a good test of the quark charges,
which it is. However, it is equivalent
to the statement that the cross section for absorbing virtual
isoscalar photons is one-ninth of the cross section for
absorbing virtual isovector photons. This  was
known to be approximately true for real photons and had been explained by
vector-meson dominance, so that it could be, and was, argued that perhaps
the 5/18 relation had
nothing to do with quarks!  Of course, the data were really
telling us that photoproduction  already suggested
non-integral quark charges.

In the same paper I discussed the then relatively well-known
fact that the experimental value of 0.18 for $\int F^{ep}_2 dx$
looked rather big compared to the parton value $\Sigma Q^2_i \langle
x_i\rangle$, even
with non-interval charges, and pointed out that  any
valence  plus uniform sea model would give a value greater than
2/9.  I then noted that this value could be `{\em easily reduced by
adding a background of neutral partons (which could be
responsible for binding the quarks)}', an idea which I recall was
attacked, in a seminar that I gave at CERN, for being not in
the spirit of the quark model!

By the end of 1969 it was known that $\sigma_L/\sigma_T = 0.2 \pm 0.2$, which
was encouraging for those of us who were proponents of the
quark model.  It would be wrong to think, however, that the
quark parton model was widely accepted at that time.  A
review talk \cite{lleweref22} that I gave in 1970 describes many other
ideas that
were still on the market (diffractive models, Harari's model,
generalized vector meson dominance, a Veneziano-like model,
etc.).
 By the end of 1970, however, it was known that the neutron and proton
structure functions were different, which laid certain ideas
(diffractive models; Harari's model) to rest, but there was
still no general consensus on the correct underlying picture.

The most important development for our field in 1971 was, of
course, 't Hooft's proof that non-Abelian gauge theories
are renormalizable.  On the theoretical deep inelastic front, I
published \cite{lleweref23} the following sum rule, derived by formal
manipulations involving the energy momentum tensor, for the
fraction of the momentum ($\epsilon$) of a high-energy nucleon carried by
 gluons:
\begin{equation}
\epsilon = 1 + \int \left({\frac{3}{4} [F^{\nu p}_2 + F_2^{\nu n}] -
\frac{9}{2} [F_2^{ep}
+ F^{en}_2]}\right) dx\;.
\end{equation}
At that time the available neutrino data could not be used to
do better than give an experimental lower bound ($\epsilon \geq 0.52 \pm
0.38)$, although it was rather obvious from the data that $\epsilon $
could not be zero.  Today the latest parton fits to the data
\cite{lleweref24}, including
all QCD corrections, give $\epsilon = 0.39$ at $q^2 = 2$, increasing to
$0.44$ at $q^2 = 20$, 0.47 at $q^2 = 2000$, and 0.48 at $q^2 = 200~000$.

Bob Jaffe has already mentioned the `light-cone algebra' of
Gell-Mann and Fritzsch, which reproduced all the general
results of the quark parton model in terms of expectation
values of bilinear light-cone operators abstracted from the
quark model \cite{lleweref25}, without admitting to the physical reality of
the quark-parton picture.  Murray Gell-Mann came to give a talk
on this at SLAC, in the auditorium in which we are meeting
today.  I was in the audience and was getting excited because I
knew (my paper was written but not published) that because he
was using a free quark model, it must imply $\epsilon = 0$, which was not
compatible with the data.  At a certain point Murray announced
that he was about to derive a sum rule using the energy
momentum tensor.  I stuck up my hand and said that I knew what
the sum rule was going to be (the sum rule above with $\epsilon = 0$)
and that it was contradicted by the data.  To this, Murray
replied {\em `You must be Llewellyn Smith, I'm pleased to meet you
but you spell your name wrongly'!}

By that time the attention of Bj and others had switched to
trying to anticipate the gross features of the final states in
deep inelastic scattering, which had not then been studied\footnote{Nor had
the gross
features of $pp$ collisions yet been established by the ISR, where
experiments began in
late 1971. It is often forgotten that the existence of a central rapidity
plateau was not
established pre-ISR, and was not universally anticipated. I remember long
discussions on
this subject in the SLAC cafeteria in 1971--72, during which David Ritson
regularly
asserted that the existence of a plateau had been disproved by cosmic ray
data which
showed that there are two fireballs!}, and other `hard' processes, such as
the production
of particles with large transverse momentum in proton collisions.  Study of
deep inelastic
scattering in field theory continued.  Gribov and Lipatov's seminal
summation of the
leading powers of ln($q^2$) in an Abelian gauge theory was published
\cite{lleweref26} in
1971, and their results were rederived using the operator product expansion
and the
renormalization group by Christ, Hasslacher and Mueller in a paper
\cite{lleweref27} that
established the techniques used later in QCD.

Asymptotic freedom and QCD  were the major theoretical
developments of 1973.  The event of the year (perhaps of the
decade) experimentally was the discovery of neutral currents by
Gargamelle.  The same experiment also produced the first $y$
distributions for charged current interactions \cite{lleweref28}, which were of
course in line with quark-parton expectations, as was implicit
in the 1972 result that $\sigma^\nu \approx 3 \sigma^{\bar\nu}$.

The first QCD calculations of corrections to scaling and to sum
rules appeared in 1974, bringing the end of the sum rule story
in sight.  This, however, was not at all clear at the time due
to the `high $y$ anomaly' in neutrino $y$ distributions measured at
Fermilab (which of course turned out to be wrong), the
observation of neutrino dimuon and trimuon events (the former
later interpreted as charm production; the latter wrong), and
above all the apparently smooth rise of $\sigma_{\bar ee}$ between 3 and
5~GeV found at
SLAC (following earlier indications that the cross-section was unexpectedly
large from
CEA).

The SLAC  $\bar ee$ data, which became public at the end of 1973, were the
focus of attention at the 1974 London conference, in a session
with 61 theoretical contributions.  In his capacity of
rapporteur B.~Richter declared \cite{lleweref29} that {\em `the data contradict
both the simple quark--parton model and the Bjorken scaling
hypothesis'}.  Commenting  on his own {\em `favourite models involving
new lepton--hadron interactions'} he remarked that, {\em `struck ... by
similar features seen in hadronic interactions'} on first seeing
the data, he had suggested that a kind of {\em `no photon
annihilation'} was involved and had  found that he was in distinguished
company (Pati and Salam).  The discovery of the $J/\Psi$ later in
1974 opened the way to understanding the $\bar ee$ data and a full
vindication of the quark-parton picture, although charmed
particles were not found until 1976.

The Ellis--Jaffe sum rule, to which I will return in the next
section, was published \cite{lleweref30} in 1974.  It should perhaps be
called an ansatz, rather than a sum rule, as it depends on an
explicit dynamical assumption.  This assumption has turned out
to be wrong, with the result that Ellis now calls it the Jaffe
sum rule and vice versa.

This ends my brief history of sum rules.  Although the
interpretation of the $J/\Psi$ as a $c\bar c$ bound state was not fully
established in 1975, and the confusion caused by the high $y$
anomaly persisted, the first indications of scaling violations
in that year encouraged proponents of QCD (it is interesting to
compare the plots of $F_2$ versus $q^2$ at fixed $x$ shown at the 1975
SLAC conference \cite{lleweref31}, which had large errors and ranged up to
$q^2$
of order $20~{\rm GeV}^2$, with the exquisitely accurate data now
available \cite{lleweref32}, which go up to $q^2$ above 10~000~GeV$^2$ at
large $x$).
The following years witnessed very convincing tests of the QCD
predictions, the development of perturbative QCD and the
anticipated \cite{lleweref33} manifestation of the gluon in three-jet events
in $e\bar e$ annihilation.

\section{Sum Rules for Spin-Dependent Lepton and Photon
Scattering}

The parton model describes deep inelastic scattering off
polarized nucleons in terms of contributions from quarks with
spins parallel $(\uparrow )$ or antiparallel $(\downarrow )$ to the spin of the
parent nucleon in the infinite momentum frame:
\begin{figure}[H]
\centering{\includegraphics[width=7cm]{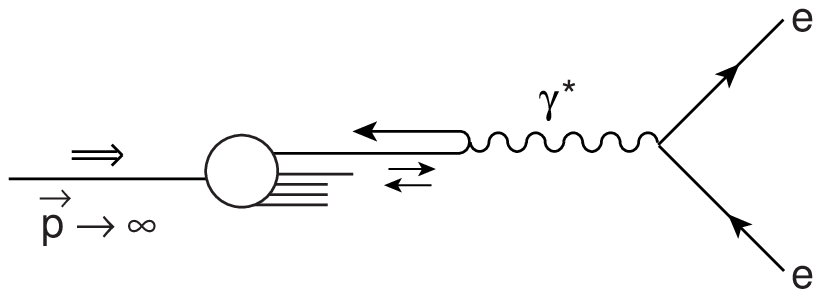}}
\end{figure}

The lowest moment of the structure function $g_1$ is related to
the numbers of such quarks thus:
\begin{equation}
\int g_1 (x) dx = \frac{1}{2} \sum Q_i^2 \left({n_i^\uparrow +
n^\uparrow_{\bar i} -
n_i^\downarrow - n^\downarrow_{\bar i}}\right)\;.
\end{equation}
It is convenient to write
$$Q^2_i = \frac{2}{9} + \frac{1}{3} I^i_3 + \frac{1}{6} Y^i$$
where $c, t$ and $b$ quarks are neglected here and in the
rest of this section. Noting that in the parton picture
\begin{equation}
\langle \lambda = 1/2, p|\bar\psi I_3 \gamma_\mu (1\mp \gamma_5) \psi
|\lambda = 1/2 ,
p\rangle = 2p_\mu\sum I_3^i \left({n_i^{\downarrow \uparrow} -
n^{\uparrow\downarrow}_{\bar i}}\right)
\end{equation}
while
\begin{equation}
\langle 1/2, p| \bar \psi I_3 \gamma_\mu\gamma_5 \psi |1/2, p\rangle = 2p_\mu
\left({-\frac{g_A}{2g_V}}\right)\;,
\end{equation}
the Bjorken sum rule, Eq.~(3), can immediately be derived.

To make predictions for the proton and neutron separately, it
is necessary to use data from hyperon decays (the $F/D$ ratio) to
obtain the $Y_i$ term and make an additional assumption to predict
the singlet piece.  Ellis and Jaffe \cite{lleweref30} assumed  that
\begin{equation}
n_s^\uparrow + n^\uparrow_{\bar s} - n_s^\downarrow - n^\downarrow_{\bar s}=0
\end{equation}
 on the grounds that there are presumably few strange
quarks in the proton and they are unlikely to be polarized.
In the parton model, this gives
$$\int g^p_i dx = 0.186\pm 0.004$$
 in serious contradiction with the observed value \cite{lleweref34} of
$0.121 \pm 0.003 \pm 0.005
\pm 0.017$ at $q^2 =5$~GeV$^2$ (the disagreement persists
when QCD
corrections  are included).

An equivalent form of the Ellis--Jaffe prediction is
\begin{equation}
\Delta \Sigma = \sum_i \left({n_i^\uparrow + n^\uparrow_{\bar i} -
n^\downarrow_i -n^\downarrow_{\bar i}}\right) = 0.58 \pm 0.03
\end{equation}
 compared to the experimental value \cite{lleweref34} of $0.19 \pm
0.05 \pm 0.04$.  The conclusion is that, on the basis of the na\"{\i}ve
parton
model, strange quarks carry a substantial fraction ($- 15\%$) of
the polarization while most ($70\%$) is not in quarks at all.
Actually the fact that the Ellis--Jaffe ansatz does not give the
na\"{\i}ve result  $\Delta  \Sigma =1$ already leads to one or both of these
conclusions without any input from deep inelastic scattering.
This also reminds us that the success of the Bj sum rule, which
relates deep inelastic data to the measured $g_A/g_V$, not to the
na\"{\i}ve three-quark model value or to some simple number derived
by counting, is highly non-trivial.

An immediate question is whether the na\"{\i}ve predictions survive
renormalization in QCD.
 Physically, it should be noted that the parton model derivation
assumes that the component of the struck quark's spin along $\vec p$
is identical to its helicity. This is a safe assumption in a
super-renormalizable theory in which the integral over the
quark's transverse momentum $\vec k_T$ converges. In a merely
renormalizable theory, however, it may not be safe to let $\vec p
\rightarrow \infty$
before integrating over $\vec k_T$. Consequently  $|\vec k_T|/|\vec p|$
cannot
necessarily be assumed small and the spin-equals-helicity assumption may not be
correct.

I shall consider this from the point of view of the operator
product expansion\footnote{Use of the operator product expansion makes it
necessary to adopt a
renormalization prescription for the relevant Wilson operators, one of
which is
afflicted by the famous anomaly in this case. The equivalent procedure of
factorizing all
the relevant Feynman diagrams into `hard' (mass singularity free) and
`soft' parts
likewise requires the adoption of a factorization prescription. In the
diagrammatic
approach, however: i) it is more apparent how to choose a prescription that
allows a
simple interpretation for the soft parts, which are normally identified
with the quark and
gluon distributions (I particularly like a prescription \cite{lleweref36}
which relates
these distributions to Fourier transforms along the light-cone of quark and
gluon
correlation functions smeared over distances $(q^2)^{-1/2}$ transverse to the
light-cone), and ii) the anomaly is never directly encountered (the
amplitudes for deep
inelastic scattering are all of course anomaly free).} in order to
facilitate comparison
with the rest of the literature.  In the case of the non-singlet
(Bjorken) sum rule, the twist-two spin-one operator, which controls the
twist-two
contribution to the Bjorken integral, is the axial isospin current.  This
current is
conserved as
$m_q\rightarrow 0$   and needs no special renormalization of its own (it is
rendered
finite by the standard QCD vertex and wave function
renormalizations).  The Bjorken sum rule and the associated
parton picture therefore survive renormalization (albeit with
the well-known correction factor $C^{NS} (q^2)
=1-\frac{\alpha_s}{\pi}\cdots$ on the
right-hand side of the sum rule).

In the singlet case the twist-two spin-one operator built from
quark fields formally has the form `$\bar\psi \gamma_\mu\gamma_5\psi$',
where the inverted
commas indicate that in this case the divergences implicit in
multiplying $\bar\psi(0)$  by $\psi (0)$ are not removed by the standard QCD
renormalizations.  Indeed, because of the anomaly, this Wilson
operator needs a special renormalization of its own and the
results depend on the renormalization scheme chosen.

It would seem natural to choose a scheme that respects QCD
gauge invariance e.g. $\overline{MS}$.  There being no gauge-invariant,
twist-two, spin-one
gluonic operator, the lowest moment of $g_1$ is controlled by the matrix
elements of the
gauge invariant ($gi$) Wilson operator (or `current' --- but this term
tempts the
na\"{\i}ve to think that it can be manipulated like a conserved current)
that I denote $[\bar \psi \gamma_\mu\gamma_5\psi]_{gi}$. The divergence of
this operator
is given by \cite{lleweref37}:
\begin{equation}
\partial^\mu [\bar \psi \gamma_\mu\gamma_5\psi]_{gi} =
\frac{\alpha_s}{2\pi} F\cdot \tilde
F + 0 (m_q)
\end{equation}
where $F$ is the QCD field tensor and $\tilde F$ its dual.  Because the
gauge invariant `current' has a `hard' divergence, its matrix
elements depend on the renormalization scale, which it is
convenient to take to be  $q^2$.  The quantity $\Delta \Sigma$ introduced
above, which enters the na\"{\i}ve parton sum rule, is therefore
replaced by $C(q^2) \Delta \Sigma(q^2)$ where
\vglue.3cm
\begin{itemize}
\item[--]  $C(q^2)$ is the coefficient
function in the operator product expansion (= hard scattering
amplitude in the language of factorized diagrams) which, as in
the non-singlet case, produces the familiar correction factor $1-
(\alpha_s/\pi)\cdots $ at finite $q^2$;
\item[--] $\Delta\Sigma (q^2)$ is given by
$$\langle p| [\bar\psi \gamma_\mu\gamma_5 \psi]_{gi}|p\rangle = 2 p_\mu
\Delta \Sigma\;.$$
\end{itemize}

The $q^2$ dependence of $\Delta \Sigma$, which arises because $q^2$ is
chosen as the
renormalization scale,  is actually quite mild. It tends
to decrease
$\Delta\Sigma$ as
$q^2$ decreases, thereby making the difference between the measured and the
Ellis--Jaffe
values bigger. A recent fit \cite{lleweref34}  gives $\Delta\Sigma (1) =
0.19 \pm 0.05 \pm 0.04$.  However, given the need for renormalization, and
that therefore
the spin projection on the $z$ axis is not equal to the helicity, it is
not obvious that $\Delta\Sigma $ should have a simple quark model
interpretation.

A number of authors \cite{lleweref40} have advocated writing
\begin{equation}
[\bar\psi \gamma_\mu\gamma_5\psi]_{gi} = [\bar\psi
\gamma_\mu\gamma_5\psi]_s + K_\mu
\end{equation}
  where $K_\mu$  is
 the gauge-dependent quantity
\begin{equation}
K_\mu = \frac{\alpha_s}{8\pi} \epsilon^{\mu\alpha\beta\gamma} A^a_\alpha
\left({F^a_{\beta\gamma}-
\frac{g}{3} f_{abc} A_\beta^b A_\gamma^c}\right)\;.
\end{equation}
As discovered long ago by Adler \cite{lleweref37} and Bardeen
\cite{lleweref41} , the
gauge-dependent current
$J_\mu^s = [\bar\psi
\gamma_\mu\gamma_5 \psi]_s$   is conserved and is the current corresponding
to the
chiral symmetry (hence the label $s$) that QCD exhibits in this limit.
Advocates of
introducing this separation say that   $K_\mu$ is {\em `obviously'} to be
identified with
the gluon contribution.  They then note that the quantity $\Delta\Sigma^q$
defined by
\begin{equation}
\langle p|J^s_\mu|p\rangle = 2 p_\mu \Delta\Sigma^q
\end{equation}
is $q^2$ independent, due to the fact that the
symmetry current is conserved as $m_q\rightarrow 0$, and claim that this is
{\em `necessary for a quark model interpretation'}.

This leads to the relation
\begin{equation}
\Delta\Sigma (q^2) = \Delta\Sigma^q - \frac{\alpha_s}{2\pi} \Delta g (q^2)
\end{equation}
where $\Delta\Sigma^q$ and $\Delta g$
are claimed to be the contributions of quarks and gluons to the
proton's spin (the gluonic term appears to vanish
asymptotically, but this is not so since $\Delta g$ varies as
$\alpha_s^{-1}$, as
is necessary to reconcile the $q^2$ dependence of $\Delta \Sigma (q^2)$ and
$\Delta\Sigma^q$).

A fit using this so-called `Adler--Bardeen scheme' gives \cite{lleweref34}
$$\Delta \Sigma^q = 0.38\pm 0.03 \pm 0.05\;,$$
 which does not help much in practice in interpreting
quantities measured in deep inelastic scattering in a simple
quark picture.  Personally, however, I am not at all convinced
that $\Delta\Sigma^q$ should have a simple interpretation.  It is true that
the divergence that appears to be responsible for spoiling the
simple parton picture resides in $K_\mu$, and that $K_\mu$ generates the
two-gluon contribution in the $t$-channel view of  $\langle p|[\bar\psi
\gamma_\mu\gamma_5\psi]_{gi} |p\rangle$.  However, $\langle p | J^s_\mu
|p\rangle$ still
contains multigluon contributions  that are not present in a na\"{\i}ve
quark picture,
and the gauge-dependent quantities $J^s_\mu$ and $K_\mu$  are  peculiar
(unphysical?)
objects, both of which necessarily couple to the unphysical `ghost' U(1)
Goldstone boson.  Furthermore, I have yet to be convinced that
the `AB scheme' used to fit the data treats all the moments of
the polarized quark and gluon distributions consistently\footnote{The fits
deal with the
distributions, not their moments. The `AB scheme' would appear to treat the
lowest $(n =0
)$ moment of the structure functions in a special way which seems to
destroy the
analyticity in $n$ that is necessary to construct distributions from moments.
Distributions can be defined directly as Fourier transforms along the
light-cone of
suitably regulated bilocal operators (see the previous footnote): from this
point of view,
the `AB scheme' and the introduction of the subtractively renormalized
quantity $K_\mu$
look very unnatural.}.

It may be an illusion to  expect
that the deep inelastic data should have a simple
interpretation in terms of the na\"{\i}ve (non field theoretical)
quark model in the singlet case in which the divergences of
field theory really matter.  In any case, more experimental
information on the structure of polarized nucleons is clearly
needed, from experiments  interpreted
with consistently defined and used quark and gluon
distributions.  In this respect, the forthcoming COMPASS
experiment at CERN, which will measure the polarization
asymmetry in events in which charm is produced (a signal that
gluons are involved), and the study of large $p_T$ production of
jets and photons in polarized $pp$ collisions at RHIC will be
particularly interesting.

New data will soon also  cast light on the interesting
question of the way in which the Drell--Hearn--Gerasimov--Iddings
(DHGI) sum rule is (presumably) satisfied and the connection
between polarization asymmetries in photoproduction and deep
inelastic scattering.  To derive the DHGI sum rule, consider
the amplitude for forward elastic scattering of a photon of
energy $\omega$ and initial/final polarization $\epsilon/\epsilon^\prime$
from a nucleon,
which may be written:
\begin{equation}
T(\omega) = \vec \epsilon^{\prime *} \cdot \vec \epsilon f (\omega) + i\vec
\sigma \cdot
(\vec
\epsilon^{\prime *} \times \vec \epsilon) g(\omega)\;.
\end{equation}
The imaginary part of the $g(\omega)$ is proportional to:
\begin{equation}
\Delta \sigma = \sigma_{\gamma N}^{1/2} - \sigma^{3/2}_{\gamma N}
\end{equation}
where $\sigma_{\gamma N}^{3/2}$   and $\sigma^{1/2}_{\gamma N}$ are
respectively the total
$\gamma N$ cross sections for the cases with total initial spins  $\pm 3/2$
and
$\pm 1/2$  along the photon direction.  DHGI noted that, on one hand,
$m^2g (\omega )/ (2\alpha\omega)$ is determined exactly by the anomalous
magnetic
moment $(\kappa)$ of the target (proton or neutron) for $\omega \rightarrow
0$, with
corrections of order $\omega^2$,
 as a result of the celebrated low energy theorem, which depends only on
Lorentz and gauge invariance.  On the other hand, assuming that $g(\omega
)$ satisfies an
unsubtracted dispersion relation, as expected on the basis of
standard considerations, this same quantity may be related to
an integral over $\Delta \sigma$, which can be expanded around $\omega = 0$
in powers
of
$\omega^2$.  Equating the two expressions gives
the DHGI sum rule:
\begin{equation}
-\frac{\kappa^2}{4} = \frac{m^2}{8\pi^2 \alpha}  \int \frac{\Delta \sigma
d\omega}{\omega}
\end{equation}
which, as noted by D--H, is {\em `of
interest because of its experimental as well as theoretical
implications'}.

A large part of the original interest of meeting what D--H
called the {\em `formidable challenge'} of testing their sum rule
experimentally, which has still not been done, was to test the
assumption of an unsubtracted dispersion relation which
(explicitly or implicitly) is also needed to derive the Adler sum
rule.  The sum rule is now generally expected to be true, and
interest has focused on the very non-trivial questions of how
it is satisfied and how the polarization and deep inelastic
asymmetries match up as  $q^2$ is varied.  I deal with these
questions in turn before finishing with some interesting
results that can be derived by applying DHGI to other processes.

Inga Karliner, in a thesis \cite{lleweref42} written at SLAC under the
supervision of Fred Gilman which singles out Sid Drell for
special thanks, constructed the contribution of $\gamma N \rightarrow \pi
N$ to the DHGI
integral using the results of phase shift analyses of
unpolarized data; she also included an estimate of the
contribution of $\gamma N \rightarrow \pi\pi N$.  Her analysis, which went
up to 1.2 GeV,
has been extended to 1.7 GeV by Workman and Arndt \cite{lleweref43} (the
changes are small).  The results (given without errors) are:
\begin{center}
\begin{tabular}{lcccc}
\hline
&$p$&$n$&$p+n$&$p-n$\\
\hline
WA
&$-257$\phantom{.5}&$-189$\phantom{.5}&$-446$\phantom{.5}&$-68$\phantom{0.5}\\
D
HGI value &$-204.5$&$-232.8$&$-437.3$&$-28.5$\\
\hline
\end{tabular}
\end{center}
Although inelastic channels are known to be important in
photoproduction, the amount by which pion
photoproduction fails to saturate  the sum rule is surprising,
especially for $p - n$, where relatively rapid convergence of the
integral is expected, given the (accidental?)
success for $p + n$.

It will be very interesting to learn from experiments
that are just starting at Jefferson Lab with CEBAF what additional channels
contribute,
including possibly  $K\Lambda^*$, $K\Sigma^*$ etc. if the failure of the
Ellis--Jaffe sum rule is due to substantial contributions from
polarized strange quarks, as well as $\rho N, \epsilon N, \omega N$ etc.  These
channels are included in a calculation by Burkert and Li \cite{lleweref44}
who estimated contributions from all $s$ channel resonances up to
and including the $F_{37}$(1950) using a mixture of experimental
input and quark model predictions.  They found for the DHGI
integral:
\begin{center}
\begin{tabular}{ccccc}
\hline
	&$p$	&$n$	&$p + n$&$	p - n$\\
\hline
 BL&	$- 203$&$	- 125$&$	- 328$&$	- 78$\\
\hline
\end{tabular}
\end{center}
The failure to reproduce the DHGI prediction (due to
non-resonant contributions? a failure of the generally
successful quark model? inconsistent  use of the quark model and
data?) heightens the interest in the forthcoming CEBAF
experiments.

These experiments will also explore the very interesting
intermediate $q^2$ region.  If we define
\begin{equation}
I(q^2) = \frac{2m^2}{|q^2|} \int g_1 (x, q^2) dx\;,
\end{equation}
then the right-hand side of the DHGI sum rule is equal to $I$ (0).  Whereas for
$p -n$ it is easy to match the large $q^2$ (Bjorken sum rule) value
of $I (q^2)$ onto the DHGI prediction for $q^2 = 0$, the situation is
very different for the proton, as shown in the following figure
\cite{lleweref45}, where
the break in the negative $I$ scale should be noted.

\begin{figure}[H]
\centering{\includegraphics[width=7cm]{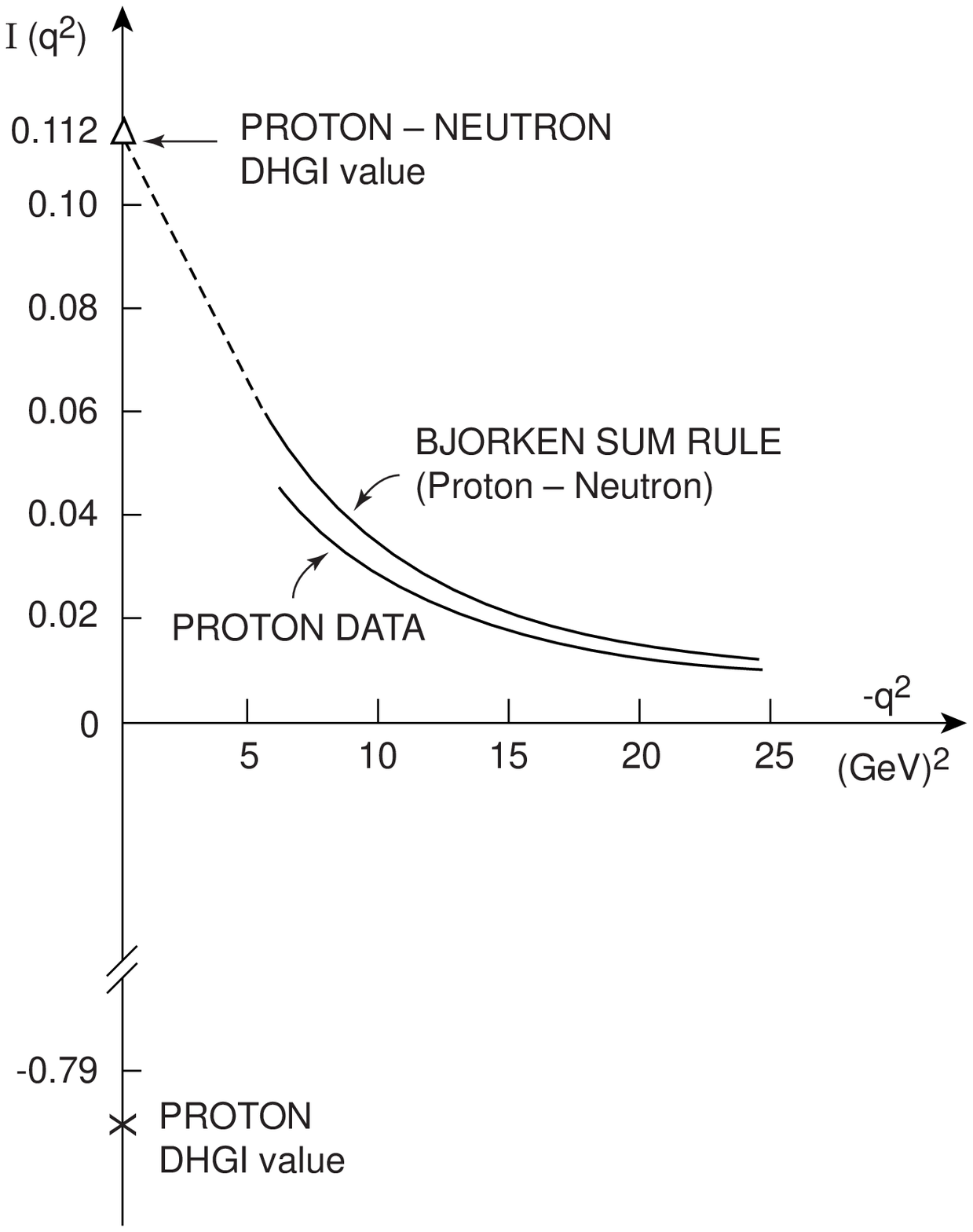}}
\end{figure}
\noindent
DHGI requires (the appropriately weighted integral of) $\Delta\sigma$ for
the proton to be
 negative for $q^2 = 0$, whereas it is positive in the deep
inelastic region. $\Delta \sigma$ is negative for photoproduction of the
dominant resonances $[\Delta , D_{13}(1520), F_{15}(1690)]$ as was explained
long ago in the quark model \cite{lleweref46}.  The $\Delta$ is produced by
an M1
transition, implying  $\sigma^{3/2}  = 3\sigma^{1/2}$.  In the harmonic
oscillator quark model, the helicity 1/2 amplitudes for
photoproducing the next two resonances are given by
\begin{eqnarray}
&&A^{\gamma p} (D_{13} ) \sim |\vec q_{cm}|^2 - \alpha^2/g\cr
&&\\
&&A^{\gamma p} (F_{15}) \sim |\vec q_{cm}|^2 - 2 \alpha^2/g\nonumber
\end{eqnarray}
where $\alpha$ is the harmonic
oscillator constant, and $g$ is  the quark $g$ factor.  It turns out
that $|\vec q_{cm}|^2_{F_{15}} \simeq 2|\vec q_{cm}|^2_{D_{13}}$   so both
amplitudes can,
`accidentally', be --- and for reasonable values of $\alpha$ and $g$ are
--- small, in
agreement with the data.

However, $|\vec q_{cm}|^2$ varies rapidly with $q^2$ and, as pointed out in
1972
by Close and Gilman \cite{lleweref47}, it is likely that $\Delta\sigma$
rapidly changes
sign as $q^2$ increases, in conformity with the deep inelastic result.
Quark model calculations by Burkert and Li \cite{lleweref44}  predict a
very complex $q^2$ behaviour (initial increase of $\Delta\sigma$, due to the
contribution of the $\Delta$, followed by a rapid decrease due to the
contributions of
other resonances), which does not seem to be consistent with
predictions based on  chiral  perturbation theory.  It will
be very interesting to see the CEBAF data.

Finally, interesting  theoretical conclusions can be derived by applying
DHGI to other targets.  In terms of $g$ factors, the sum rule
reads
\begin{equation}
-(g-2)^2=\frac{m^2}{8\pi\alpha} \int \frac{d\omega}{\omega} \Delta \sigma\;.
\end{equation}
 In pure QED, $\Delta\sigma$ is of order $\alpha^2$ (e.g. for an electron
target the
lowest-order process is  $\gamma e \rightarrow \gamma e$).  It follows that
\begin{equation}
 g-2 =0 + 0 (\alpha)
\end{equation}
 i.e. consistency requires that $g = 2$ to lowest order in QED, a
result which also follows from other considerations such as
requiring good high-energy behaviour  (this argument could only
fail if the DHGI integral required a subtraction, which however
would introduce a new arbitrary constant and spoil
renormalizability).  Plugging (31) back into (30) then gives:
\begin{equation}
\int \frac{d\omega}{\omega} \Delta \sigma_{QED} = 0 (\alpha^3)
\end{equation}
i.e. the integral of the order $\alpha^2$ contribution to $\Delta \sigma$ must
vanish, as pointed out by Altarelli, Cabbibo and Maiani \cite{lleweref48}.
This
highly non-trivial constraint, which provides a check of order
$\alpha^2$ calculations, was generalized by Brodsky and Schmidt
\cite{lleweref49} using a
loop counting argument to:
\begin{equation}
\int \frac{d\omega}{\omega} \Delta \sigma^{tree}_{\gamma a\rightarrow bc} =
0\;.
\end{equation}
This also provides non-trivial checks on calculations, and a
potential diagnostic for new physics e.g. by checking whether
the sum rule is satisfied by   cross-sections such as $\gamma e\rightarrow
W\nu$ that
could be measured at a linear collider.

Brodsky and Drell have pointed out  \cite{lleweref50} that if a lepton
$(L)$ had
substructure which could be photo-excited above a threshold $m^*$
there would be an additional order $\alpha$ term:
\begin{equation}
\int_{m^*} \Delta\sigma_{\gamma L\rightarrow x} \frac{d\omega}{\omega}
\end{equation}
on the right-hand side of (32).  It follows that:
$$g-2 = 0 + 0 (\alpha^2) + 0 \left({\frac{m_L}{m^*}}\right)\;.
$$

Brodsky and Drell's paper is mainly devoted to showing how this interesting
relation arises in explicit models.  On the one hand, it allows us
to interpret possible deviations from the prediction of QED for
$g -2$  in terms of a mass scale ($m^*$) for substructure.  On
the other hand, it assures us that $g = 2$ would emerge naturally
in any composite model in which $m^*$ is extremely large, although
no one knows how to make a model in which the mass of a lepton
could be very small compared to the mass of its constituents.

\section{Concluding Remark}

It is pleasing to end with this nice result of Sid's.  His paper
with Stan Brodsky  nicely illustrates the pragmatic and realistic
style of Sid's many contributions to our field.  General
arguments are used, but  only when they can be substantiated
and illustrated using specific models, and relevance to
possible experiments is paramount.  I would like to end by
thanking Sid for what he has given to our field, and to me
personally, not only in terms of physics, but also by  promoting this style
of doing
physics as the long-time leader of the outstanding SLAC theory group.


\newpage

\begin{thebibliography}{99}
\bibitem{lleweref1} S.L. Adler, Phys. Rev. {\bf 143} (1966) 1144.
\bibitem{lleweref2} J.D. Bjorken, Phys. Rev. {\bf 148} (1966) 1467;\\
Phys. Rev. Lett. {\bf 16} (1966) 408.
\bibitem{lleweref3} J.M. Cornwall and R.E. Norton, Phys. Rev. {\bf 177}
(1969) 2584.
\bibitem{lleweref4} SMC collaboration, D. Adams et al., Phys. Lett.
{\bf B396} (1997) 338 and Phys. Rev. {\bf D56} (1997) 5330;
B. Adeva et al., Phys. Lett. {\bf B412} (1997) 414 and Phys. Rev. {\bf
D58} (1998) 112001.
\bibitem{lleweref5} S.D. Drell and A.C. Hearn, Phys. Rev. Lett. {\bf 16}
(1966) 908.
\bibitem{lleweref6} S.B. Gerasimov, Yad. Fiz. {\bf 2} (1965) 839 [Sov. J.
Nucl. Phys. {\bf
2} (1966) 598].
\bibitem{lleweref7} C.K. Iddings, Phys. Rev. {\bf 138} (1965) B446.
\bibitem{lleweref8} J.D. Bjorken, Proc. 1967 International Symposium on
Electron and
Photon Interactions at High Energy, IUPAP: CONF-670923.
\bibitem{lleweref9} J.D. Bjorken, in
Proc. 1967 Varenna
Summer School, Academic Press (1968).
\bibitem{lleweref10} J.D. Bjorken, Phys. Rev. {\bf 179} (1969) 1547.
\bibitem{lleweref11} W.K.H. Panofsky, in Proc. XIV International
Conference on High Energy
Physics, Eds.~J.~Prentki \& J.~Steinberger, CERN (1968), p.~23.
\bibitem{lleweref12} K. Gottfried, Phys. Rev. Lett. {\bf 18} (1967) 1174.
\bibitem{lleweref13} S.D. Drell and C. Schwartz, Phys. Rev. {\bf 112}
(1958) 568.
\bibitem{lleweref14} 
NMC Collaboration, M. Aneodo et al., Phys. Rev. {\bf
D50} (1994) R1.
\bibitem{lleweref15} J.D. Bjorken and E.A.~Paschos, Phys. Rev. {\bf 185}
(1969) 1975;
{\bf D1} (1970) 3151.
\bibitem{lleweref16} S.D. Drell,  D.J. Levy and T.M. Yan, Phys. Rev.  Lett.
{\bf 22}
(1969) 744;
Phys. Rev. {\bf 187} (1969) 2159;
Phys. Rev. {\bf D1} (1970) 1035;
Phys. Rev. {\bf D1} (1970) 1617; \\
S.D. Drell and T.M. Yan, Phys. Rev. {\bf D1} (1970) 2402;
Phys. Rev. Lett. {\bf 24} (1970) 855;
Phys. Rev. Lett. {\bf 23} (1970) 316 (E:25, 902); Ann.  Phys. (N.Y.) {\bf
66} (1971)
555.
\bibitem{lleweref17}C.G.~Callan and D. Gross, Phys. Rev. Lett. {\bf 22}
(1969) 136.
\bibitem{lleweref18} D. Gross and  C.H. Llewellyn Smith, Nucl. Phys. {\bf
B14} (1969) 337.
\bibitem{lleweref19} J.H. Kim, D.A. Harris et al., Phys. Rev. Lett. {\bf
81} (1998) 3595.
\bibitem{lleweref20} R.~Jackiw and G.~Preparata, Phys. Rev. Lett. {\bf 22}
(1969)
975;\\ S.L.~Adler and W.K.~Tung, Phys. Rev. Lett. {\bf 22} (1969) 978.
 \bibitem{lleweref21} C.H. Llewellyn Smith, Nucl. Phys. {\bf B17} (1970) 277.
\bibitem{lleweref22} C.H. Llewellyn Smith, CERN TH - 1188 (1970).
\bibitem{lleweref23} C.H. Llewellyn Smith, Phys. Rev. {\bf D4} (1971) 2392.
\bibitem{lleweref24} R.G. Roberts, private communication.
\bibitem{lleweref25} H. Fritzsch, M. Gell-Mann and H. Leutwyler,
Phys. Lett. {\bf 47B} (1973) 365.
\bibitem{lleweref26} V.N. Gribov and L.N.~Lipatov,  Sov. J. Nucl. Phys.
{\bf 15} (1972)
438.
\bibitem{lleweref27} N. Christ, B. Hasslacher and A.H. Mueller, Phys. Rev. {\bf
D6} (1972) 3543.
\bibitem{lleweref28} D.H. Perkins, in Proc. 5th Hawaii Topical Conference
(1973),
University of Hawaii Press (1974).
\bibitem{lleweref29} B. Richter, in Proc. XVII International Conference on
High Energy
Physics, Science Research Council, Rutherford Laboratory (1974), p.~IV--37.
\bibitem{lleweref30} J. Ellis and R.L. Jaffe, Phys. Rev. {\bf D9} (1974) 1444;
{\bf D10} (1974) 1669.
\bibitem{lleweref31} Proc. 1975 International Symposium on Lepton and Photon
Interactions at High Energy, Stanford University (1975).
\bibitem{lleweref32} T. Doyle, invited talk at  1998 ICHEP, Vancouver.
\bibitem{lleweref33} J. Ellis, M.K. Gaillard and G. Ross, Nucl. Phys. {\bf
B111} (1976) 253.
\bibitem{lleweref34} SMC collaboration, B. Adeva et al., Phys. Rev. {\bf
D58} (1998) 112002.
\bibitem{lleweref36} C.H. Llewellyn Smith, in `Symmetry Violations in
Subatomic Physics',
p.~139, Eds. B.~Castel and P.J. O'Donnell, Proc. 1988 CAP--NSERC Summer
Institute, World
Scientific (1989).
\bibitem{lleweref37} J.~Bell and R. Jackiw, Nuovo Cimento {\bf 60A} (1969)
47;\\
R.~Jackiw, in `Lectures on Current Algebra and its Applications', Eds.
S.B.~Treiman,
R.~Jackiw and D.J.~Gross, Princeton University Press (1972);\\ S.L. Adler,
Phys. Rev. {\bf
177} (1969) 2426 and in
`Lectures on Elementary Particles and Quantum Field Theory',
Eds.~S.~Deser, M.~Grisaru and
H.~Pendleton, MIT Press (1971), Vol.~1, p. 1.
\bibitem{lleweref40} A.V. Efremov and O.V. Teryaev, in `Hadron
Interactions',
Eds. J. Fischer, P. Kolar and V. Kundrat, Czech. Acad. Sci. Inst. Phys.
(1988) p. 302.
G. Altarelli and G.G. Ross, Phys. Lett. {\bf B212} (1988) 391;
R.D. Carlitz, J.C. Collins and A.H. Mueller, Phys. Lett. {\bf B214} (1988)
229.
\bibitem{lleweref41} W.A. Bardeen, Nucl. Phys. {\bf B75} (1975) 246.
\bibitem{lleweref42} I. Karliner, SLAC--180 (1974); Phys. Rev.~{\bf D7}
(1973) 2717.
\bibitem{lleweref43} R.L. Workman and R.A. Arndt, Phys. Rev. {\bf D45}
(1992) 1789.
\bibitem{lleweref44} V. Burkert and Z. Li, Phys. Rev. {\bf D47} (1993) 46.
\bibitem{lleweref45} N. De Botton, in `High Energy Spin Physics',
Eds.~K-H. Althoff \&
W.~Meyer, Springer Verlag (1991), p.~419.
\bibitem{lleweref46} L.A. Copley, G. Karl and E. Obryk, Phys. Lett.
{\bf 29B} (1969) 117 and Nucl. Phys. {\bf B13} (1969) 303
\bibitem{lleweref47} F.E. Close and F.J. Gilman, Phys. Lett. {\bf 38B}
(1972) 541.
\bibitem{lleweref48} G.~Altarelli, N. Cabibbo and L.~Maiani, Phys. Lett.
{\bf 40B} (1972)
415.
\bibitem{lleweref49} S.J. Brodsky and I. Schmidt, Phys. Lett. {\bf B351}
(1995) 344.
\bibitem{lleweref50} S.J. Brodsky and S.D. Drell, Phys. Rev. {\bf D22}
(1980) 2236.
\end{thebibliography}
\end{document}